\documentclass[twocolumn,prl,aps,floats,floatfix,superscriptaddress,amsmath,amssymb,nofootinbib,longbibliography]{revtex4-1}

\usepackage[final]{graphicx}
\usepackage{hyperref}
\usepackage{amsmath}
\usepackage{amsfonts}
\usepackage{amssymb}
\usepackage[english]{babel}
\usepackage{float}
\usepackage{xcolor}
\usepackage{physics}
\usepackage{bm}
\usepackage{booktabs}
\usepackage{siunitx}

\usepackage{amsmath}
\usepackage{accents}
\newlength{\dhatheight}
\newcommand{\doublehat}[1]{%
    \settoheight{\dhatheight}{\ensuremath{\hat{#1}}}%
    \addtolength{\dhatheight}{-0.35ex}%
    \hat{\vphantom{\rule{1pt}{\dhatheight}}%
    \smash{\hat{#1}}}}

\begin{document}

\title{First Experimental Limit on the Thermal Solar Neutrino Flux}

\author{Cecilia Ferrari}
\email{ferraric@mit.edu}
\affiliation{Laboratory for Nuclear Science, Massachusetts Institute of Technology, Cambridge, MA 02139, USA}
\author{Gonzalo Herrera}
\email{gonzaloh@mit.edu}
\affiliation{Kavli Institute for Astrophysics and Space Research, Massachusetts Institute of Technology, Cambridge, MA 02139, USA}
\affiliation{Laboratory for Particle Physics and Cosmology, Harvard University, Cambridge, MA 02138, USA}
\author{Brooke Russell}
\email{russell3@mit.edu}
\affiliation{Laboratory for Nuclear Science, Massachusetts Institute of Technology, Cambridge, MA 02139, USA}

\date{\today}

\begin{abstract}
The neutrino sky below 165\,keV is yet to be explored. 
This region provides a unique probe of stellar cooling mechanisms through the detection of thermal solar neutrinos and the low-energy tail of the $pp$ solar cycle. 
Here, we investigate prospects for probing this regime via neutrino capture on tritium. 
Analyzing KATRIN public data, we set the first experimental bound on the thermal solar neutrino flux $\Phi/\Phi_{\mathrm{SSM}} < 1.86 \times 10^{18}$ at 95\%~CL ($1.58\times10^{18}$ at 90\%~CL), and show that a $100\;\text{kg}\cdot\text{yr}$ exposure would constrain the thermal solar neutrino component to $\Phi/\Phi_{\mathrm{SSM}} \lesssim 10^4$ and detect the low-energy $pp$ flux at the Standard Solar Model (SSM) level. 
Neutrino--electron elastic scattering from $pp$ cycle neutrinos are identified as an irreducible background for neutrino capture searches.
\end{abstract}

\maketitle

\emph{Introduction.}---With a peak flux density of $\sim$10$^9\,$cm$^{-2}\,$s$^{-1}$\,MeV$^{-1}$~\cite{Haxton:2012wfz}, the thermal neutrino flux dominates the neutrino sky in the spectral range bookended by the standard cosmic neutrino background (C$\nu$B) and the low-energy tail of solar fusion~\cite{RevModPhys.92.045006}.
Thermal neutrino emission is a critical cooling mechanism in the evolution of red giants, pre-supernovae, white dwarfs, and neutron stars~\cite{Winget_2004, 10.1111/j.1365-2966.2007.12342.x}.
Although thermal neutrinos contribute negligibly to our sun's evolution, solar thermal neutrino detection provides an independent cross check on the solar core temperature~\cite{HAXTON2000263} and chemical composition~\cite{Vitagliano2017}, two keystones of the standard solar model (SSM)~\cite{Serenelli:2011py}.
Furthermore, thermal neutrinos may hold key insights into nonstandard neutrino electromagnetic properties~\cite{PhysRevD.98.016009}.

Thermal neutrinos are thusly dubbed given their production origin through a variety of thermal mechanisms --- pair production, photo-processes, plasma processes, bremsstrahlung, recombination,~\cite{1996ApJS..102..411I} and stimulated $\nu\bar{\nu}$ emission~\cite{PhysRevD.98.016009}.
With energies $\lesssim$10\,keV, the thermal neutrino flux is well below the detection threshold of contemporary neutrino experiments.
For context, the Borexino experiment has measured the lowest-energy neutrinos detected to date --- $pp$ neutrinos above 165\,keV measured through elastic neutrino-electron scattering~\cite{2014Natur.512..383B}.

In this Letter, we investigate for the first time the feasibility of detecting sub-$100$~keV solar neutrinos (thermal neutrinos and the low-energy portion of the $pp$ spectrum) through neutrino capture on tritium, $\nu_e + {}^3\mathrm{H} \to {}^3\mathrm{He} + e^-$, the reaction proposed for C$\nu$B detection~\cite{PhysRev.128.1457,Cocco2007,Lazauskas_2008,PhysRevD.77.113014,LI2010261,Faessler:2011qj}. 
Unlike the monochromatic endpoint signature of the C$\nu$B, solar neutrinos in this regime imprint a continuous electron spectrum from the tritium beta decay endpoint to tens of keV above it.

This paper is organized as follows.
We derive the first upper limit on the thermal solar neutrino flux using KATRIN publicly released data. 
Next, we discuss the rate and spectral shape of elastic neutrino-electron scattering processes, which constitute an irreducible background in tritium-based neutrino capture searches.
Finally, we derive projected upper limits for the thermal solar neutrino flux and the low-energy tail of the $pp$ flux through neutrino capture searches.
The quantification of the relative impact of the experimental parameters --- exposure, dynamic range above the tritium beta decay endpoint, and the role of the irreducible neutrino-electron elastic scattering background --- is emphasized.

\emph{Low-Energy Solar $\nu$ Capture on Tritium Nuclei}\label{sec:capture}---Neutrino capture on monoatomic tritium,
$\nu_e + {}^3\mathrm{H} \to e^- + {}^3\mathrm{He}$,
produces an electron with kinetic energy
$K_e = Q_\beta + E_\nu$, where
$Q_\beta = 18.591~\mathrm{keV}$~\cite{LNHB_H3}.
For a differential neutrino flux $\Phi(E_\nu)$, the electron kinetic energy ($T$) spectrum from capture on $N_T$ tritium atoms over an exposure time $\tau$ is
\begin{equation}
\frac{dN}{dT}
=
N_T\,\sigma(E_\nu)\,\Phi(E_\nu)\,\tau\,.
\label{eq:capture_spectrum}
\end{equation}
The total capture cross section~\cite{Cocco2007,Behrens1971,Behrens1982} is
\begin{equation}
\sigma(E_\nu) =
\frac{G_F^2 \cos^2\!\theta_C}{\pi}\,(1+3g_A^2)\,p_e E_e F(Z,E_e)\,,
\label{eq:sigma_capture}
\end{equation}
where
$G_F$ is the Fermi constant,
$\cos\theta_C = V_{ud}$ is an element of the Cabibbo-Kobayashi-Maskawa (CKM) matrix,
$g_A$ is the axial-vector weak coupling constant of the nucleon,
and $p_e$ and $E_e$ are the outgoing electron momentum and total energy, respectively. 
The Coulomb interaction with the ${}^3\mathrm{He}$ daughter nucleus is included through the relativistic Fermi function~\cite{Behrens1971,Behrens1982}
\begin{equation}
F(Z,E_e)=
2(1+\gamma)\,(2p_eR)^{2(\gamma-1)}\,e^{\pi\eta}
\frac{|\Gamma(\gamma+i\eta)|^2}{|\Gamma(2\gamma+1)|^2}\,,
\label{eq:fermi_function}
\end{equation}
with
$\gamma=\sqrt{1-(\alpha Z)^2}$,
$\eta=\alpha Z E_e/p_e$,
and
$R=1.2\,A^{1/3}~\mathrm{fm}$, for $A=3$.

\begin{figure}
\centering
\includegraphics[width=\columnwidth]{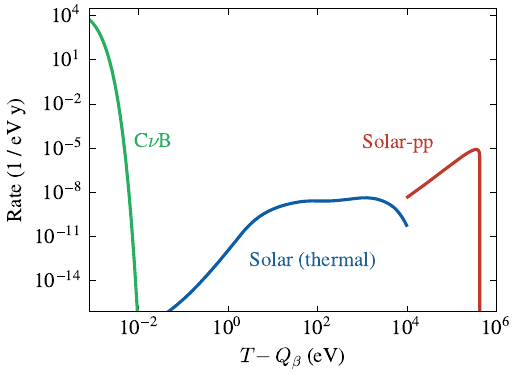}
\caption{Kinetic energy ($T$) differential capture rate spectra for solar thermal (solid blue), solar $pp$ (solid red), and C$\nu$B (solid green) neutrinos on a 100~g tritium target mass assuming the SSM flux normalization. %Solar $pp$ neutrino-electron ($\nu{e}$) elastic scattering off target source atomic electrons is an intrinsic background to low-energy solar neutrino capture searches. 
%The solar $pp$ neutrino-electron ($\nu{e}$ scattering final state electron kinetic energy spectrum (dashed gray) is also shown.
}
\label{fig:thermal_capture}
\end{figure}

We compute the differential neutrino capture rate on tritium using the Standard Solar Model (SSM) thermal neutrino flux from Ref.~\cite{Vitagliano2017}\footnote{Flavor mixing is ignored in the computation of the solar $\nu_e$ flux at Earth in Ref.~\cite{Vitagliano2017}.}.
Solar thermal and solar $pp$ differential capture rate spectra are shown in Figure~\ref{fig:thermal_capture} for a reference $100~\mathrm{g}$ tritium target mass.
The C$\nu$B capture rate for an equivalent target mass is shown for comparison.

%Figure~\ref{fig:thermal_capture} additionally reports the intrinsic background to these searches due to electrons emitted in solar $pp$ neutrino elastic scattering processes on the atomic tritium target. These electrons are characterized by an approximately flat kinetic energy spectrum ranging from \SI{1}{eV} up to about \SI{200}{keV}, where the solar $pp$ neutrino spectrum cuts off. A description of this background is given in the following Section. 

\emph{KATRIN-derived Thermal Solar $\nu$ Limit.}---The KATRIN experiment performs precision integral spectroscopy measurements of the bi-molecular tritium beta decay endpoint to directly measure the effective electron antineutrino mass~\cite{drexlin2026katrinexperiment}.
We use KATRIN public data from~\cite{KATRIN2025} to derive the first experimental upper limit on the thermal solar neutrinos flux. 

The thermal solar neutrino flux normalization
%differential rate of thermal solar neutrino captures $R$ as a function of KATRIN spectrometer retarding energy $qU$ 
is found by fitting the spectral region corresponding to a spectrometer retarding energy $qU > 18576$~eV, 
% above the bi-molecular tritium beta decay endpoint
with a modified form of the KATRIN response function~\cite{KATRIN2025}: 
\begin{equation}
\small
R(qU)_{i} = A \frac{m_{i}}{m_{1}}\int_{qU}^{E_{\rm max}} R_\text{diff}(E)\, \bar{f}(E - qU)\, dE   + R_\text{bg,i} + R_\text{s}(qU)\,,
\label{eq:fit_model}
\end{equation}
where normalization parameter $A$ encapsulates detector efficiencies and the thermal neutrino flux normalization, $m_{i}$ is the source mass, $m_{1}$ is the tritium source mass in the first data-set campaign, $R_\text{diff}(E)$ is the differential rate of the thermal solar neutrino capture signal computed for a reference 100\,g tritium source at the Standard Solar Model (SSM) flux~\cite{Vitagliano2017} and a bi-molecular tritium endpoint of 18575.0~eV (see Figure~\ref{fig:thermal_capture}), $\bar{f}$ is the electron transfer function (approximated here as a step function), $R_\text{s}(qU)$ combines the fixed KATRIN background rate contributions (referred to as slope and Penning backgrounds in~\cite{KATRIN2025}), and $R_{{\rm bg},i}$ is a data-set dependent constant background contribution (accounting for other radioactive and cosmic backgrounds as described in~\cite{KATRIN2025}).
The integral upper limit  $E_{\rm max}=(x-10)$~keV reflects the KATRIN focal plane detector ROI selection cut, where 10~keV is the energy gained by the electrons from the post-filter acceleration and $x$ is a dataset dependent energy selection upper bound~\cite{KATRIN2025}. 
The label $i$ denotes a specific dataset among the seven measurement campaigns.
$A$ and $R_{{\rm bg},i}$ were free parameters in the fit.
Consistent with Ref.~\cite{Aker_2022}, the fit exclusively considers decays resulting in the molecular ground-state. 

\begin{figure}[t]
\centering
\includegraphics[width=\columnwidth]{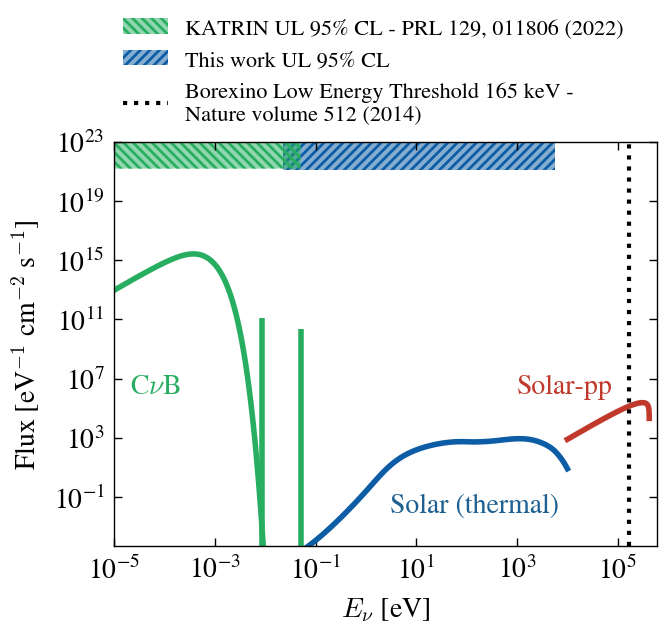}
\caption{Upper limits on the neutrino flux as a function of energy. The green hatched region shows the existing KATRIN C$\nu$B upper limit at 95\% CL~\cite{KATRINcnub2022}. The blue hatched region shows the thermal solar neutrino upper limit derived in this work at 95\% CL. The solid curves show the expected C$\nu$B (green) solar thermal neutrino (blue) and $pp$ (red) fluxes from Standard Model predictions. The vertical black dotted line shows the lowest energy value of detected neutrinos set by Borexino in Ref.~\protect\cite{2014Natur.512..383B}.}
\label{fig:flux_exclusion}
\end{figure}

A Bayesian fit was performed using the python-based probabilistic programming framework \textit{PyMC}~\cite{pymc2023}, maximizing a combined least squares likelihood with uniform priors.
A 90\% CL upper limit on the $A$ normalization parameter of $5.11 \times 10^{10}$ was deduced. 
The total efficiency $\varepsilon$ is expressed as
\begin{equation}
    \varepsilon = \varepsilon_{\rm transport}\varepsilon_{\rm geom}\varepsilon_{\rm FSD},
\label{eq:tot_eff}
\end{equation}
where $\varepsilon_{\rm transport}$ is the total transport efficiency (taken as $0.5$), $\varepsilon_{\rm geom}=0.18$~\cite{Aker_2022} is the geometrical acceptance, and $\varepsilon_{\rm FSD}=0.43$~\cite{Aker_2022} is the molecular ground-state fraction of the final-state distribution. 
Accounting for the ratio of the 100\,g reference mass to the $m_{1}$ first data campaign normalization mass, we find an upper limit on the solar thermal neutrino flux of
${\Phi}/{\Phi_\text{SSM}} < 1.58 \times 10^{18}$ at 90\% CL,
where $\Phi_{\rm SSM}$ is the solar thermal flux central prediction from Ref. \cite{Vitagliano2017}.
At 95\% CL, we obtain an upper limit of $1.86 \times 10^{18}$.
Our upper limit on the thermal solar neutrino flux covers the energy range from $\sim 0.1$~eV to $\sim 10$~keV, as shown in Figure~\ref{fig:flux_exclusion}. 
This limit is the first direct experimental probe of the neutrino sky in this energy regime.

We highlight that parameterizing $\bar{f}$ with a step function was a conservative choice.
A more realistic transfer function, taking into account the energy-dependent transmission probability~\cite{KATRIN2025}, results in a factor of two increase in the flux.
%Similarly, the flux increases by 15\% is gained once, instead of assuming the monoatomic tritium Q-value, $18575.0$~eV is used. 

Due to their sub-dominant impact, contributing to the total KATRIN experimental background ($\mathcal{O}(10^{-1})$) with less than $\mathcal{O}(10^{-11})$ counts per second, solar $pp$ contributions are omitted in this analysis.

\emph{$\nu_{e}$ Scattering Irreducible Background.}\label{sec:elastic_scatterings}---Neutrino--electron elastic scattering, $\nu + e^- \to \nu + e^-$, of solar $pp$ neutrinos off of atomic electrons poses a potential irreducible background for any neutrino capture search above the tritium beta-decay endpoint.
These interactions produce electrons with a continuous kinetic energy that extends up to
\begin{equation}
T_\text{max} = \frac{2 E_\nu^2}{m_e + 2 E_\nu}\,,
\label{eq:Tmax}
\end{equation}
where $E_\nu$ is the incoming neutrino energy and $m_e$ is the electron mass. 
Scattered electrons from $pp$ $\nu_{e}$ scattering have a  $T_\text{max} \approx 261$~keV, well above the beta decay endpoint.

\begin{figure}
\centering
\includegraphics[width=\columnwidth]{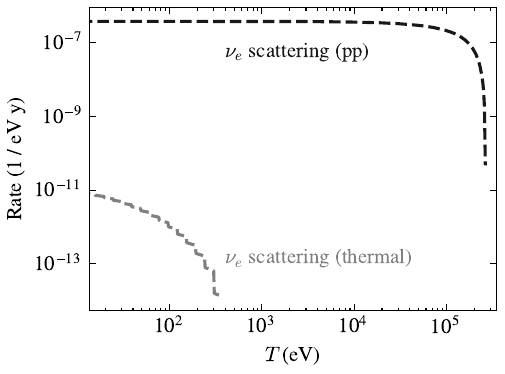}
\caption{Solar thermal (light gray) and $pp$ (dark gray) neutrinos elastic scattering on tritium electrons as a function of the emitted electron kinetic energy $T$. Rate is computed for a 100~g tritium source at the SSM flux normalization. The effect of atomic binding energy (13.6~eV) limits the extension of the spectra to lower energies.}
\label{fig:rate_comparison}
\end{figure}

To quantify the solar $pp$ $\nu_{e}$ scattered electron kinetic energy spectrum, we convolve the differential cross section $\frac{d\sigma}{dT}(E_\nu,T)$ with the SSM $pp$ neutrino spectrum $\Phi_\text{pp}(E_\nu)$, taking into account the three flavor components:
\begin{equation}
\frac{dR_\text{scat}}{dT} = N_e \sum_{i}^{3}\int_0^{E_\nu^\text{max}} \!\!\Phi_{\text{pp},i}(E_\nu)\,\frac{d\sigma_{i}}{dT}(E_\nu, T)\,dE_\nu\,,
\label{eq:scatter_rate}
\end{equation}
where $N_e$ is the number of target electrons in the source and ${E_\nu^\text{max}}$ is the maximum neutrino energy from the $pp$ spectrum. 
The differential cross section for $\nu_e$--$e$ scattering in the Standard Model~\cite{Vogel:1989iv} is
\begin{equation}
\frac{d\sigma}{dT} = \frac{2\, G_F^2\, m_e}{\pi}\left[ g_L^2 + g_R^2 \left(1 - \frac{T}{E_\nu}\right)^{\!2} - g_L\, g_R\, \frac{m_e\, T}{E_\nu^2} \right],
\label{eq:dsigma_dT}
\end{equation}
where $T$ is the electron kinetic energy, $g_L = \tfrac{1}{2} + \sin^2\!\theta_W$ and $g_R = \sin^2\!\theta_W$ are the left- and right-handed couplings for $\nu_e$ (including both charged- and neutral-current contributions), and $\theta_W$ is the weak mixing angle.
In the energy regime near the tritium Q-value, the electron kinetic energy ($T \approx Q_\beta = 18.591$~keV) is much smaller than the typical $pp$ neutrino energy ($E_\nu \sim 100$--$420$~keV), returning $T/E_\nu \ll 1$. In this limit, the differential cross section~(\ref{eq:dsigma_dT}) is dominated by the constant term $g_L^2$, with negligible dependence on $T$. Similarly, we compute the electron spectrum from thermal solar neutrino–electron scattering on tritium. Figure~\ref{fig:rate_comparison} reports these two spectra.

%Over the energy region of interest of the KATRIN experiment~\cite{KATRIN2025}, covering up to $\mathcal{O}(100)$~eV above the tritium Q-value, a spectral variation below 0.05\%, is foreseen. More generally, given an region of interest of width $\Delta T$ centered at $T_0 = Q_\beta$, the relative variation of $d\sigma/dT$ scales as
%\begin{equation}
%\frac{\Delta(d\sigma/dT)}{d\sigma/dT} \sim \frac{g_R^2}{g_L^2 + g_R^2}\,\frac{\Delta T}{E_\nu} \ll 1\,.
%\end{equation}

Solar $pp$ $\nu_{e}$ scattering constitutes an irreducible background for the proposed C$\nu$B searches~\cite{Betti_2019}, with a limited impact of $\mathcal{O}(10^{-7})$ counts for a \SI{1}{eV} window per year.
 Near $Q_{\beta}$ energies
%, it is also worth mentioning that atomic binding effects are negligible \cite{Chen:2016eab}, and 
the ionizing contribution from coherent elastic neutrino nucleus scattering (CE$\nu$NS) \cite{Freedman:1973yd, Billard:2014yka, Blanco-Mas:2024ale, DeRomeri:2024iaw, AtzoriCorona:2025gyz} and the associated Migdal effect \cite{Ibe:2017yqa, Bell:2019egg, Herrera:2023xun, Maity:2024hzb} are expected to be negligible compared to the elastic neutrino-electron scattering rate.

\emph{Sub-100 keV Solar Neutrino Sensitivity Projection.}---To determine the experimental requirements for reaching the solar neutrino flux predicted by the SSM, we compute projected 90\% CL upper limits for a counting experiment on the neutrino flux ratio $\Phi/\Phi_{\mathrm{SSM}}$ as a function of $W$, representing the counting integration window upper bound subtracted from the tritium Q-value. 
For a given exposure (source mass $m$ and exposure time $\tau$) and detection efficiency $\varepsilon$, the expected number of signal counts in an energy window $[Q_\beta,\, Q_\beta + W]$ is
\begin{equation}
s(W) = \frac{\Phi}{\Phi_{\mathrm{SSM}}} \times \varepsilon \times \frac{m \tau}{100\;\text{g}} \times \int_{Q_\beta}^{Q_\beta + W} \frac{dR}{dE_e}\, dE_e \,,
\label{eq:signal_counts}
\end{equation}
where $dR/dE_e$ is the differential capture rate per unit energy per unit time for a 100\,g tritium source at the SSM flux, computed for both thermal and solar $pp$ neutrinos from the cross sections described in Fig. \ref{fig:rate_comparison}. 
The detection efficiency $\varepsilon$ is consistent with Eq.~\ref{eq:tot_eff}.

Assuming that $\nu{e}$ scattering dominates the background budget, the expected background counts in the window $[Q_\beta,\, Q_\beta + W]$ are
\begin{equation}
\small
b(W) = \frac{m \tau}{100\;\text{g}} \times N_e \times \int_{Q_\beta}^{Q_\beta + W} dT \sum_{i}^{3}\int_{E_\nu^{\min}(T)}^{E_\nu^{\max}} \frac{d\Phi_{\mathrm{pp},i}}{dE_\nu}\,\frac{d\sigma_{i}}{dT}\, dE_\nu \,,
\label{eq:bkg_counts}
\end{equation}
where $N_e$ is the number of electrons in 100\,g of tritium and $E_\nu^{\min}(T) = \frac{1}{2}(T + \sqrt{T(T+2m_e)})$ is the minimum neutrino energy that can produce an electron of kinetic energy $T$. The differential $pp$ neutrino flux carries a 55$\%$ fraction of electron neutrinos, with the remaining 45$\%$ in muon/tau neutrinos. This has a mild effect on the elastic neutrino-electron scattering cross section of $\sim 35\%$.

Upper limits on $\Phi/\Phi_{\text{SSM}}$ are derived via profile likelihood ratio. We generate the null distribution as $n \sim \text{Poisson}(b)$ and 
compute $q(n) = -2\ln[\mathcal{L}(\mu=0|\hat{\theta})/\mathcal{L}(\mu|\doublehat{{\theta}})]$ 
for each $n$. From the toy-simulated $q$-distribution under the signal hypothesis, the $p$-value is computed as the tail probability of null 
$q$ in this distribution. The 90\% CL limit $\mu_{\text{UL}}$ is found imposing 
$P(p\text{-value}|\mu)=0.10$. For finite backgrounds, we report the median 
expected limit $\mu_{\text{UL}}(b) = \sum_n P(n|b) \, \mu_{\text{UL},n}(n,b)$, 
which averages over observations under the null hypothesis. 
In the background-free limit, we use 
$\mu_{\text{UL}} = -\ln(0.10) \approx 2.3$ counts.
In both cases, the normalization flux limit 
is obtained via $\Phi_{\text{UL}}/\Phi_{\text{SSM}} = \mu_{\text{UL}} / [s \varepsilon (m/100{\rm g}) \tau]$.

%We derived upper limits using Poisson statistics at 90\% confidence level. In the background-free regime, the 90\% CL upper limit on the mean signal is $\mu_{90} = -\ln(0.10) \simeq 2.3$ counts, corresponding to the case where zero events are observed with zero expected background. In the presence of the elastic scattering background, we compute the median expected upper limit by averaging over Poisson-distributed observations 
%we set the observed count to the median background expectation, $n_{\mathrm{obs}} = \lfloor b + 0.5\rfloor$, and determined the signal upper limit $\mu_{90}$ by solving \cite{Feldman:1997qc}
%\begin{equation}
%\sum_{n}^{n_{\mathrm{obs}}} \frac{(\mu_{90} + b)^n\, e^{-(\mu_{90}+b)}}{n!} = 0.10 \,.
%\label{eq:poisson_ul}
%\end{equation}
%The flux ratio upper limit is then obtained as
%\begin{equation}
%\left.\frac{\Phi}{\Phi_{\mathrm{SSM}}}\right|_{90\%} = \frac{\mu_{90}}{s(W)\big|_{\Phi/\Phi_{\mathrm{SSM}}=1}} \,.
%\label{eq:flux_ul}
%\end{equation}

\begin{figure}
\centering
\includegraphics[width=\columnwidth]{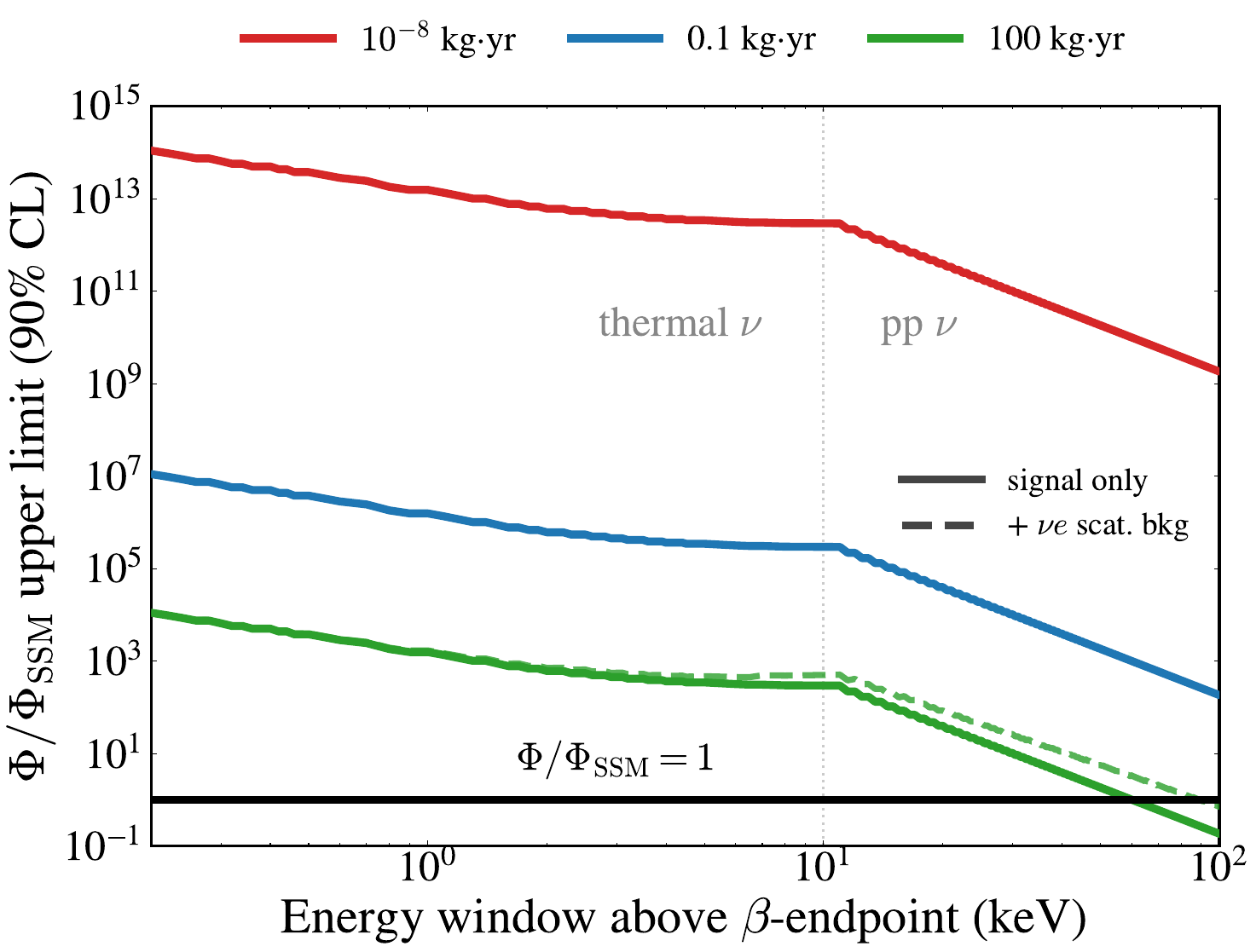}
\caption{Projected 90\% CL upper limits on the solar neutrino flux ratio $\Phi/\Phi_{\mathrm{SSM}}$ as a function of the energy window above the tritium $\beta$-endpoint. The red curve shows the limit obtained for a KATRIN-like exposure and a background free experiment. The blue and green curves show projections for $0.1\;\text{kg}\cdot\text{yr}$ and $100\;\text{kg}\cdot\text{yr}$ exposures, respectively, with solid lines for the signal-only (background-free) case and dashed lines including the irreducible $\nu_e$--$e$ elastic scattering background. The horizontal black line marks $\Phi/\Phi_{\mathrm{SSM}}=1$. The vertical dotted grey line marks the transition from a thermal neutrino-dominated capture rate to a low-energy pp-neutrino tail-dominated one.}
\label{fig:sensitivity_final}
\end{figure}

The resulting sensitivity curves are presented in Figure~\ref{fig:sensitivity_final}. Three exposure scenarios are compared: a KATRIN-like exposure ($10\;\mu\text{g}\cdot\text{yr}$), a $0.1\;\text{kg}\cdot\text{yr}$ scenario, and a $100\;\text{kg}\cdot\text{yr}$ scenario. 
For each projected exposure, solid lines show the background-free sensitivity while dashed lines include the $\nu{e}$ scattering background. 
This scattering background has a negligible impact at moderate exposures ($\lesssim 1\;\text{kg}\cdot\text{yr}$), where the expected number of both signal and background counts remains well below unity and the Poisson upper limit is insensitive to the background level. 
At larger exposures, however, the integrated elastic scattering rate---which exceeds the capture rate for energy windows below $\sim 80$~keV---accumulates enough counts to degrade the sensitivity by up to a factor of $\sim 4$ at $W = 100$\,keV for the $100\;\text{kg}\cdot\text{yr}$ scenario. 
The background-free $100\;\text{kg}\cdot\text{yr}$ projection reaches $\Phi/\Phi_{\mathrm{SSM}} < 1$ for energy windows above $\sim 40$\,keV, indicating that an exposure of this order would be required to detect solar $pp$ neutrinos via tritium capture above the $\beta$-decay endpoint. 
When the elastic scattering background is included, this detection threshold shifts to wider energy windows ($\sim 80$\,keV), underscoring the importance of background characterization for next-generation experiments approaching SSM sensitivity.

\iffalse
\begin{figure}
\centering
\includegraphics[width=\columnwidth]{sensitivity_final.pdf}
\caption{Projected 90\% CL upper limits on the solar neutrino flux ratio $\Phi/\Phi_{\mathrm{SSM}}$ as a function of the energy window above the tritium $\beta$-endpoint. The gray curve shows the KATRIN result anchored at $W=200$\,eV. The dark blue and cyan curves show projections for $0.1\;\text{kg}\cdot\text{yr}$ and $10\;\text{kg}\cdot\text{yr}$ exposures, respectively, with solid lines for the signal-only (background-free) case and dashed lines including the irreducible $\nu_e$--$e$ elastic scattering background. The horizontal black line marks $\Phi/\Phi_{\mathrm{SSM}}=1$.}
\label{fig:sensitivity_final_2}
\end{figure}
\fi

%\include{conclusion}  %=======================================================
\emph{Conclusions.}---We have presented the first experimental upper limit on the thermal solar
neutrino flux using KATRIN 2025 data,
$\Phi/\Phi_{\mathrm{SSM}} < 1.86 \times 10^{18}$ at 95\%~CL ($1.58\times 10^{18}$ at 90\%~CL), bridging
the gap between the lowest-energy neutrinos ever detected---solar $pp$
neutrinos at $E \gtrsim 165$~keV by Borexino---and the cosmic neutrino
background upper limit established by KATRIN~\cite{PhysRevLett.129.011806}. 
Our sensitivity projections show that a $100\;\text{kg}\cdot\text{yr}$ tritium exposure with an energy window extending $\sim 100$~keV above the $\beta$-decay endpoint would reach the SSM $pp$ neutrino flux, while constraining the thermal component to $\Phi/\Phi_{\mathrm{SSM}} \lesssim 10^4$. 
We identify neutrino--electron elastic scattering as an irreducible background that degrades the capture sensitivity by up to a factor $\sim 4$ at the largest exposures, confirming that the thermal solar neutrino flux at the SSM level remains beyond the reach of near-term tritium-based searches.  
Nevertheless, this technique opens a path toward measuring the low-energy
tail of the solar $pp$ spectrum and constraining exotic neutrino physics
at energy scales previously inaccessible to direct detection.

%The capture signatures studied here are complementary to probes of solar thermal neutrinos through electron scattering in liquid xenon detectors, which are explored in a concurrent companion study \cite{ArguellesEtAlInPreparation}. Crucially, capture searches provide a clean test of the low-energy tail of the $pp$ flux, negligible for neutrino-electron elastic scattering signatures.

While the limits obtained in this work remain far from a detection of the thermal solar neutrino flux and the low-energy tail of the $pp$ flux, it is useful to recall that sensitivity gains of this scale have occurred before in astroparticle physics. 
For example, the first WIMP dark matter direct detection experiment to ever run, about 40 years ago \cite{Ahlen:1987mn}, set a constraint on WIMPs that was about $\sim 11$ orders of magnitude weaker than current limits. Comparable progress has also been achieved in other rare event searches. Early double beta decay experiments probed half-lives at the $\sim 10^{15}$--$10^{17}\,\mathrm{yr}$ scale, whereas present neutrinoless double beta decay searches reach beyond $10^{26}\,\mathrm{yr}$ \cite{Inghram:1949qu, KamLAND-Zen:2022tow}. 
Similarly, the first proton decay searches constrained nucleon lifetimes at the $\sim 10^{20}$--$10^{22}\,\mathrm{yr}$ level, while Super-Kamiokande now sets limits above $10^{34}\,\mathrm{yr}$ in benchmark channels \cite{PhysRev.96.1157, Super-Kamiokande:2020wjk}. 
Improvement in several-decade physics sensitivity has astroparticle physics precedent.
We remain hopeful that the low-energy neutrino sky will eventually be mapped.

\begin{acknowledgments}
\emph{Acknowledgments.}---CF and BR are supported by the Massachusetts Institute of Technology Department of Physics. The work of GH was supported by the Neutrino Theory Network Fellowship with contract number 726844.
\end{acknowledgments}

\emph{Data Availability.}---The KATRIN publicly available data release~\cite{KATRIN2025} was used in this manuscript. There is no publicly available software supporting this manuscript. Requests for further information should be sent to the authors.

\bibliographystyle{apsrev4-1}
\renewcommand{\bibsection}{\section*{\refname}}
\bibliography{references}

\appendix

\section{End Matter}

%The expected number of capture events per year at the SSM flux normalization is obtained by integrating Eq.~\eqref{eq:capture_spectrum} over the relevant energy range:
%\begin{equation}
%N_\text{events} = \int_{Q_\beta}^{Q_\beta + E_\nu^\text{max}} \frac{dN}{dE_e}\, dE_e\,.
%\end{equation}
%For the KATRIN source activity of $A_\text{KATRIN} = 23.9$~GBq (corresponding to $N_T \approx 3.8 \times 10^{18}$ atoms), we find $N_\text{th} \approx 2.4 \times 10^{-11}$ thermal capture events per year and $N_\text{pp} \approx 3.1 \times 10^{-6}$ pp capture events per year. For a PTOLEMY-scale experiment with 100~g of tritium, these numbers become $N_\text{th} \approx 1.2 \times 10^{-5}$ and $N_\text{pp} \approx 1.6$~events per year.

\emph{Theoretical uncertainties on neutrino capture rate.}---The dominant theoretical uncertainty on the thermal solar capture rate arises from the SSM thermal neutrino flux prediction, which carries an uncertainty of $\sim 10\%$ set by the core temperature and electron/ion density profiles~\cite{Vinyoles2017}. Additional sub-dominant uncertainties, also affecting the solar $pp$ capture spectrum, include:
\begin{itemize}
\item The axial coupling constant $g_A = 1.2764 \pm 0.0013$, contributing a $\sim 0.2\%$ uncertainty on $\sigma$.
\item The Fermi function $F(Z, E_e)$, which is known to sub-percent precision for the relevant energies; radiative corrections are at the $\sim 0.1\%$ level.
\item The CKM element $V_{ud} = 0.97420 \pm 0.00021$, contributing a $\sim 0.04\%$ uncertainty.
\item Final-state distributions (FSD) of the daughter ${}^3\text{HeT}^+$ molecule. Only the ground-state fraction ($\sim 43\%$ of decays) produces electrons at the nominal $E_e = Q_\beta + E_\nu$; excited final states shift the electron energy downward by up to $\sim 100$~eV, broadening the signal. We account for this via the efficiency factor $\epsilon_\text{FSD} \approx 0.43$ in the flux conversion.
\end{itemize}

In total, the theoretical prediction for the capture rate is robust at the $\sim 10\%$ level, dominated by the SSM flux uncertainty, which is far below the current experimental sensitivity.

%\subsection{Solar thermal neutrinos scattering off atomic tritium electrons}
%\label{subsec:elastic_scatterings_th}

%Following the methodology described in Section~\ref{sec:elastic_scatterings}, we compute the electron spectrum from thermal solar neutrino–electron scattering on tritium. The results are shown in Figure~\ref{fig:rate_comparison}.

%\begin{figure}
%\centering
%\includegraphics[width=\columnwidth]{fig/th_e_andpp_scattering_rate.pdf}
%\caption{Solar thermal (light gray) and $pp$ (dark gray) neutrinos elastic scattering on tritium electrons as a function of the emitted electron kinetic energy $T$. Rate is computed for a 100~g tritium source at the SSM flux normalization. The effect of atomic binding energy (13.6~eV) limits the extension of the spectra to lower energies.}
%\label{fig:rate_comparison}
%\end{figure}

\emph{Analysis Methods with the KATRIN Open Dataset.}---KATRIN data are collected to optimize the antineutrino mass sensitivity, scanning the energy region in the neighborhood of the tritium beta decay end point. 
The 2025 KATRIN data release featured five data campaigns with a total experimental lifetime of 259 days~\cite{KATRIN2025}. 
The five data taking campaigns are subdivided into seven data-sets labeled as KNM1, KNM2, KNM3-NAP, KNM3-SAP, KNM4-NOM, KNM4-OPT and KNM5, according to the detector configuration. 
For each of these data-sets ($i$), the KATRIN fit model can be summarized as:
\begin{equation}
\small
R(qU)_{i} = A_{T}R_{\beta,\rm int}   + R_\text{bg,i} + R_\text{s}(qU)\, + R_{RW}(E),
\label{eq:fit_model_katrin}
\end{equation}
where $R_\text{s}(qU)$ includes the slope and Penning contributions, $R_\text{bg,i}$ is the constant background for the data-set $i$, $R_{RW}(E)$ is the rear-wall spectrum background, due to tritium compounds formed with detector materials, and $R_{\beta,\rm int}$ is the integral of the differential spectrum of tritium beta decay:
\begin{equation}
\small
R_{\beta,\rm int} = \int_{qU}^{E_{\mathrm{max},i}} R_\text{T,diff}(E)\, {f}(E - qU)\, dE 
\label{eq:RintKATRIN}
\end{equation}
where $E_{\mathrm{max},i}$ is the KATRIN focal plane detector ROI selection cut, $R_{\rm T,diff}$ is the differential bi-molecular tritium beta spectrum and $f$ is the detector transfer function described in~\cite{KATRIN2025}. 
For analysis simplicity, we omitted the division of some of the datasets into subsets, called {\it patches}; dataset subsets were re-merged, taking into account the relative efficiencies, to reduce the number of correlated free parameters. 

As discussed in the main text,  in order to derive an upper limit on the thermal neutrino flux, we restrict our analysis to the data points of each data-set above the beta decay end-point of the rear-wall spectrum.
With this selection cut, we can exclude the $R_{RW}$ component from our fit model.
Thus, the only free parameters in our fit model (Equation~\ref{eq:fit_model}) are $A$, the thermal solar neutrino spectrum normalization, common to all data-sets, and the dataset dependent background $R_{bg, i}$.

\begin{figure}
\centering
\includegraphics[width=\columnwidth]{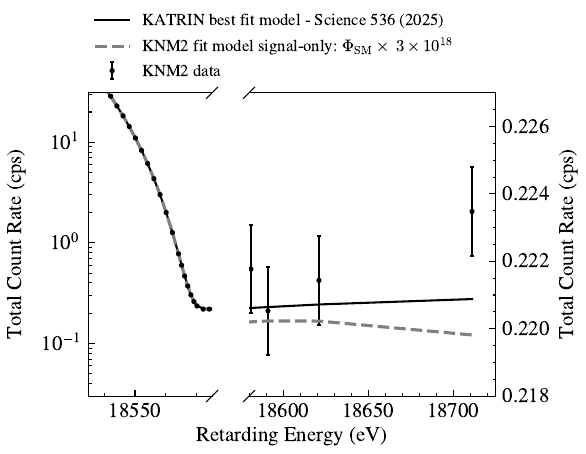}
\caption{KNM2 dataset (black errorbars) reported along the Ref~\protect\cite{KATRIN2025} KATRIN best fit model (solid-black compared with our modified model (dashed-gray), where we set $R_{\rm bg, KNM2}$ to zero and we used a normalization for the thermal neutrino flux corresponding to $3\times10^{18}$ times the SSM prediction.}
\label{fig:KATRIN_2_phi1e10_corrected}
\end{figure}

As an example, Figure~\ref{fig:KATRIN_2_phi1e10_corrected} shows the KNM2 dataset along with the best fit model of Ref.~\cite{KATRIN2025} in addition to our fit, including the thermal solar neutrino spectrum, with a normalization of $3\times10^{18}$ and simultaneously setting the $R_{\rm bg, KNM2}$ to zero.

A noted departure from Equation~\ref{eq:RintKATRIN}, we use a Heavy-side transfer function $\bar{f}$ in Equation~\ref{eq:fit_model}; as a consequence we omit detector nuisance parameters in the fit. 
We checked the effect of using $\bar{f}$ with respect to a more realistic transfer function \textit{a posteriori} and verified our upper limit was conservative by more than a factor $2$ for both 90\% and 95\% CL upper limits.

Given the choice of merging the KATRIN dataset subsets, we use a combined negative-2 log-likelihood computed as the sum over the seven datasets of their relative Least Square likelihood:
\begin{equation}
    -2\log(\mathcal{L}) = \sum_{i}\sum_{k}\left(\frac{C^{\rm obs}_{i,k}-R(qU_{k})_{i}}{\sigma^{{\rm obs}}_{i,k}}\right)^2\, ,
\label{eq:likelihood}
\end{equation}
where $C^{\rm obs}_{i,k}$ are the observed counts in dataset $i$ at set potential $qU_{k}$, $\sigma^{{\rm obs}}_{i,k}$ the corresponding uncertainty and $R(qU_{k})_{i}$ is the model reported in Equation~\ref{eq:fit_model}.

The fit was performed using a Bayesian technique withthe \textit{PyMC} python package~\cite{pymc2023},  utilizing the NUTS sampler~\cite{hoffman2011nouturnsampleradaptivelysetting}, in order to safely include in the upper limit estimation the possible parameters correlations. To even further reduce the correlation effects between the $R_{bg,i}$ parameters and $A$ parameter, a more hierarchical representation basis was used. Specifically, we decided to recast the $R_{bg,i}$ parameters as:
\begin{align}
    &R_{bg,{\rm KNM3-SAP}} = R_{bg,{\rm KNM3-SAP}}\\
    &R_{bg,j} = R_{bg,{\rm KNM3-SAP}} + R_{{\rm offset},j}
\end{align}
where $R_{bg,{\rm KNM3-SAP}}$ and $R_{{\rm offset},j}$ were used as free parameters in the fit with $j$ ranging in [KNM1, KNM2, KNM3-NAP, KNM4-NOM, KNM4-OPT, KNM5]. This choice was dictated by the fact that the KNM3-SAP data-set featured the lowest of the background rates among the considered data-sets.

The prior distributions for the parameters were chosen to be agnostic. To this end, we used for each parameter a uniform distribution:
\begin{align}
    &A = \mathrm{Uniform}(0, 10^{16}),\\
    &R_{bg,{\rm KNM3-SAP}} = \mathrm{Uniform}(0, 1),\\
    &R_{{\rm offset},j} =  \mathrm{Uniform}(0.0015, 0.3).
\end{align}
Bounds were chosen to be broad enough to as to not incur into over-constraining parameters  or posteriors clipping. 
This was later verified by performing the fit with different bounds values, returning an $A$ upper limit always compatible within percent level with our nominal result.

From the posterior of the $A$ parameter we extract the relative upper limit at 90\%, 95\% and 99\% CL by computing the 90, 95 and 99 percentile of the distribution. 
To translate the $A$ upper limits into thermal neutrino flux normalization ones, we scale for the acceptances, namely the total electron transport efficiency (conservatively assumed to be 0.5), the geometrical acceptance (0.18~\cite{Aker_2022}) and the fraction to ground state (0.43~\cite{Aker_2022}), and normalize for the KATRIN tritium source with respect to the one assumed in the signal rate computation (\SI{100}{g}) as reported in Figure~\ref{fig:thermal_capture}.

For systematic checks, we also performed run-by-run fits. The results of this check are reported in Table~\ref{tab:run-by-run-fits}.
\begin{table}[]
\caption{Results in terms of upper limits at 90 and 95 \% CL when the analysis is run per single KATRIN data-set, compared to those resulting from the combined analysis.}
    \label{tab:run-by-run-fits}
    \centering
    \begin{tabular}{llll}
        Data-set Label & Live Time (h) & UL 90\% CL &UL  95\% CL\\
        \hline
         KNM1 &  522 & $2.67\times10^{19}$ & $2.85\times10^{19}$\\
         KNM2 & 694 & $2.44\times10^{18}$ & $3.07\times10^{19}$\\
         KNM3-NAP & 220 & $8.69\times10^{18}$ & $9.29\times10^{18}$\\ 
         KNM3-SAP &  224 & $2.53\times10^{18}$ & $2.70\times10^{18}$\\ 
         KNM4-NOM &  835 & $2.69\times10^{18}$ & $2.89\times10^{18}$\\ 
         KNM4-OPT & 432 & $2.77\times10^{18}$ & $2.94\times10^{18}$\\
         KNM5 &  1226 & $2.92\times10^{18}$ & $3.13\times10^{18}$\\
         \hline
         Combined & 4153 & $1.58\times10^{18}$ & $1.86\times10^{18}$\\
         \hline
    \end{tabular}
\end{table}
We observe that the combined analysis of the full $259$ days KATRIN exposure returns an upper limit improved by a factor $10$ with respect to the ones obtained by analyzing only the KNM1 dataset $21.75$ days. This improvement is not only given by the larger accumulated exposure but mainly due to the larger $qU$ set point probed in successive datasets and the great efforts done by the KATRIN collaboration to reduce the overall background in between different runs. It is worth noticing, in this regard, that for example taking into consideration KNM4-OPT and KNM3-SAP, the first featuring a double live time with respect to the second, the set upper limits for the thermal neutrino flux normalization are comparable. This result is due to the more extended energy region probed in the latter with respect to the first campaign. 
% this is very badly written, but I think it could be a nice starting point for more qualitative analysis

We  also performed a frequentest fit, by always exploiting the likelihood of Eq.~\ref{eq:likelihood}, via iminuit package~\cite{iminuit} and found slightly more conservative upper limits when the $A$ free parameter was unconstrained to be positive. 
Table~\ref{tab:freq_fit} reports the results of the frequentist approach. 

\begin{table}
    \centering
    \caption{Results for iminuit unconstrained fit considering likelihood in Eq.~\ref{eq:likelihood}. From the profiled likelihood in the signal normalization parameter $A$ and the efficiency and mass scale conversion factors described in the text an upper limit of $2.69\times10^{18}$ at 90\% CL ($3.20\times10^{18}$ at 95\% CL) is found on the thermal solar neutrino flux normalization.}
    \label{tab:freq_fit}
    \begin{tabular}{lll}
         Parameter &  Mean & STD \\
         \hline
         $A$	& $-3\times 10^{10}$ & $4\times 10^{10}$\\
         $R_{\rm bg,KNM1}$ & 0.304 & 0.014 \\
         $R_{\rm bg,KNM2}$ & 0.27 & 0.05	 \\
         $R_{\rm bg,KNM3-NAP}$ & 0.251 & 0.032  \\
         $R_{\rm bg,KNM3-SAP}$ & 0.17& 0.06  \\
         $R_{\rm bg,KNM4-NOM}$ & 0.18 & 0.06 \\
         $R_{\rm bg,KNM4-OPT}$ & 0.18 & 0.06  \\
         $R_{\rm bg,KNM5}$ & 0.19 & 0.06  \\
         \hline
    \end{tabular}
\end{table}

\end{document}